\documentclass{iosart2c}
\usepackage{setspace}  

\usepackage[T1]{fontenc}
\usepackage{times}%

\usepackage{dcolumn}
\usepackage{graphics}
\usepackage{graphicx}
\usepackage{adjustbox}
\usepackage{multirow}
\usepackage{booktabs}
\renewcommand{\arraystretch}{0.8}
\usepackage{listings} 
\usepackage{relsize}
\usepackage{amsmath}
\usepackage[T1]{fontenc}
\usepackage{times}%
\usepackage{amsmath}
\usepackage[draft]{hyperref}

\usepackage{etoolbox}
\AtBeginEnvironment{align}{\setcounter{subeqn}{0}}
\newcounter{subeqn} \renewcommand{\thesubeqn}{\theequation\alph{subeqn}}%
\newcommand{\subeqn}{%
 ~\refstepcounter{subeqn}
  \tag{\thesubeqn}
}

\usepackage{dcolumn}
\usepackage{textcomp}
\usepackage{mathtools}

\usepackage{cite} 
\usepackage{enumitem}	
\setlist{parsep=0pt,listparindent=\parindent}

\usepackage{algpseudocode}
\usepackage{algorithmicx}
\usepackage{algorithm}

\usepackage{xcolor}

\definecolor{myviolet}{rgb}{0.5, 0, 0.8}
\definecolor{mygreen}{RGB}{10,150,10}
\definecolor{myblue}{rgb}{0,0,1}
\algdef{SE}[DOWHILE]{Do}{doWhile}{\algorithmicdo}[1]{\algorithmicwhile\ #1}
\newcolumntype{d}[1]{D{.}{.}{#1}}

\begin{document}

\begin{frontmatter}                           

\title{ParaLarH: Parallel FPGA Router based upon Lagrange Heuristics}

\runningtitle{ParaLarH}


\author[A]{\fnms{Rohit} \snm{Agrawal}},
\author[A]{\fnms{Kapil} \snm{Ahuja}\thanks{Corresponding author. E-mail: kahuja@iiti.ac.in.}},
\author[A]{\fnms{Dhaarna} \snm{Maheshwari}}
and
\author[B]{\fnms{Akash} \snm{Kumar}}
\runningauthor{R. Agrawal, K. Ahuja, D. Maheshwari and A. Kumar}
\address[A]{Data \& Computational Sciences Lab, Indian Institute of Technology Indore, Simrol 453552, India.
}
\address[B]{Center for Advancing Electronics, Technische Universit\"at Dresden, 01062 Dresden, Germany. 
}

\begin{abstract}
Routing of the nets in Field Programmable Gate Array (FPGA) design flow is one of the most time consuming steps. Although Versatile Place and Route (VPR), which is a commonly used algorithm for this purpose, routes effectively, it is slow in execution. One way to accelerate this design flow is to use parallelization. Since VPR is intrinsically sequential, a set of parallel algorithms have been recently proposed for this purpose (ParaLaR and ParaLarPD).

These algorithms formulate the routing process as a Linear Program (LP) and solve it using the Lagrange relaxation, the sub-gradient method, and the Steiner tree algorithm. Out of the many metrics available to check the effectiveness of routing, ParaLarPD, which is an improved version of ParaLaR, suffers from large violations in the constraints of the LP problem (which is related to the minimum channel width metric) as well as an easily measurable critical path delay metric that can be improved further.

In this paper, we introduce a set of novel Lagrange heuristics that improve the Lagrange relaxation process. When tested on the MCNC benchmark circuits, on an average, this leads to halving of the constraints violation, up to 10\% improvement in the minimum channel width, and up to 8\% reduction in the critical path delay as obtained from ParaLarPD. We term our new algorithm as ParaLarH. Due to the increased work in the Lagrange relaxation process, as compared to ParaLarPD, ParaLarH does slightly deteriorate the speedup obtained because of parallelization, however, this aspect is easily compensated by using more number of threads.
\end{abstract}

\begin{keyword}
FPGA routing\sep Linear programming\sep Sub-gradient method\sep  Lagrange relaxation\sep Lagrange heuristic
\end{keyword}

\end{frontmatter}
\section{Introduction}\label{Introduction}

The Electronic Design Automation (EDA) process has been the single biggest factor behind the thriving of the semiconductor industry in the last fifty years. However, it is very time consuming with routing taking a big percentage of this time. In this paper, we focus on a large subset of this problem, i.e. the expensive Field Programmable Gate Array (FPGA) \cite{JIFS_FPGA} routing process. FPGA routing is computationally expensive because the common standard algorithm to perform routing, i.e. Versatile Place and Route (VPR \cite{VPR}) is intrinsically slow. One way to accelerate routing is to exploit parallelization capabilities of the modern High Performance Computing (HPC) machines. Since VPR is fundamentally sequential, new parallel routing algorithms need to be developed. 

One of the first attempts in parallelizing this routing process was done in \cite{ParaLaR}. Here, the authors formulated the problem as a Binary Integer Linear Program (BILP), applied the Lagrange relaxation to eliminate constraints, and then solved the resulting optimization problem using the sub-gradient method and the Steiner tree algorithm. The final algorithm was termed as ParaLaR. 

In one of our recent works \cite{ParaLarPD}, we substantially improved the constraints violation drawback of ParaLaR. We achieved this by developing a more problem specific version of the sub-gradient method and fine tuning the size of its iterative step. The final algorithm was termed as ParaLarPD. When tested on the MCNC benchmark circuits, on an average, ParaLarPD achieved about a 20\% reduction in the constraints violation (that is related to the metric of the minimum channel width) as compared to ParaLaR. There are two other important metrics used to determine the efficiency of the FPGA routing process; the total wire length and the critical path delay, with the later one being more easily measurable. In \cite{ParaLarPD}, we showed that on an average, ParaLarPD achieved the same total wire length but a slightly deteriorated critical path delay when compared with ParaLaR.

In this work, we sizably improve ParaLarPD further. Since the core problem of this class of parallel routing algorithms is the constraints violation, we design a family of Lagrange heuristics to improve the Lagrange relaxation process. The new algorithm is defined as ParaLarH. When tested on the MCNC benchmark circuits, on an average, our ParaLarH halves the constraints violation when compared with ParaLarPD, which is the \textit{first} improvement. \textit{Second}, on an average, 10\% reduction is achieved in the related minimum channel width metric. The average total wire length obtained by ParaLarH is almost the same as that obtained by ParaLarPD. The \textit{third} improvement obtained by ParaLarH, when compared to ParaLarPD, is in the average critical path delay (8\% reduction). 

The use of heuristics to reduce the constraints violation in ParaLarH does incur an overhead such that the total routing time is slightly increased (when compared to ParaLarPD). However, this drawback is easily fixable by using more threads in a parallel setting. To further demonstrate the usefulness of our approach, we also compare ParaLarH with the original algorithm that ParaLarPD improved upon (ParaLaR \cite{ParaLaR}), the common standard algorithm used for routing (VPR \cite{VPR}), and two other algorithms that are sparingly prevalent (RVPack \cite{YuanWangThesis} and GGAPack2 \cite{YuanWangThesis}). ParaLarH gives results that are overall best among these as well.

The rest of this paper has four more sections. In Section \ref{Background}, we present the ParaLarPD algorithm from \cite{ParaLarPD}. Our Lagrange heuristic, its variants, and the resulting algorithm of ParaLarH are discussed in Section \ref{Proposed LH}. In Section \ref{Experimental Results}, we present the experimental results. Finally, conclusions and future work are given in Section \ref{Conclusion}.

\section{Background}\label{Background}

The routing problem in FPGA or a electronic circuit is formulated as a weighted grid graph $G(V,E)$, where \textit{V} and \textit{E} are the sets of certain vertices and edges, respectively, and there is a cost associated with each edge \cite{ParaLaR, ParaLarPD}. In this grid graph, we have three types of vertices; the net vertices, the Steiner vertices, and the other vertices. A net is represented as a~set $N \subseteq V$ consisting of net vertices with other types of vertices playing a supporting role.

Here, the goal is to find a~route for each net such that the union of all the routes will minimize the total path cost of the graph \textit{G}, which is directly proportional to the total wire length of FPGA. To achieve this objective, the problem of routing of nets is formulated as an LP problem given by~\cite{ParaLaR} (ParaLaR paper).
\begin{align}
& \min_{x_{e,i}} \sum_{i=1}^{N_{nets}}\sum_{e\in E}w_{e}x_{e,i}, \label{eq:LP}  \\
& \text{{Subject to }} A_{i}x_{i}=b_{i}, i=1,2,...,N_{nets}, ~\refstepcounter{equation}\label{eq:LP_c1} \subeqn \\ 
& \qquad\qquad  x_{e,i}=0 \hspace{0.1cm}or \hspace{0.1cm} 1, \text{{and}} \label{eq:LP_c2}\subeqn \\ 
&\qquad\qquad  \sum_{i=1}^{N_{nets}}x_{e,i}\le W, \forall e\in E \label{eq:LP_c3} \subeqn
\end{align}
with meaning of each variable given in Table \ref{tab_LPVariables}. The equality constraints guarantee that a valid route is formed for each net (these are implicitly satisfied by our solution). The inequality constraints are the channel width constraints that restrict the number of nets utilizing an edge to \textit{W}. These constraints also relate to our other complementary requirement, that is, the minimization of the channel width of each edge (achieved by an iterative reduction in the solution process).
\begin{table}[!h]
\caption{Summary of the symbols with their meanings as used in LP \eqref{eq:LP}-\eqref{eq:LP_c3}.}\label{tab_LPVariables}
\centering
\small	
\setlength{\tabcolsep}{2pt}
{\begin{tabular}{ll}\toprule
Symbols & Meaning  \\
\midrule
$x_{e,i}$ & The binary decision variables that can have value \\
	      & either 0 (if net \textit{i} does not utilize an edge \textit{e}) or 1  \\
	      & (if net \textit{i} utilizes an edge \textit{e}) \\
$N_{nets}$ & The number of nets  \\
\textit{E} & The set of edges with \textit{e} denoting one such edge \\
$w_e$ & The cost/ time delay associated with the edge \textit{e} \\
\textit{W} & A constant (input and iteratively reduced) \\ 
$A_{i}$ & The node-arch incidence matrix (the constraints  \\
	    &   matrix of the minimum cost flow problem) \\
$x_{i}$ & The vector of all $x_{e,i}$ that represents the route of \\ 
	    & the $i^{th}$ net\\
$b_{i}$ & The demand/ supply vector, which signifies the \\ 
	    & amount of cost flow to the $i^{th}$ net \\
\bottomrule
\end{tabular}}{}
\end{table}

The inequality constraints need to be relaxed or eliminated. This is because they introduce dependencies between the routing of different nets leading to the difficulty in solving the LP in a parallel manner. The Lagrange relaxation~\cite{Lagrangian} is a technique where the constraints can be eliminated by integrating them into the objective function. This introduces Lagrange multipliers $\lambda_{e}$ for each constraint, with relaxation carried out by adding $\lambda_{e}$ times the corresponding constraint to the objective function. That is, instead of the LP given in \eqref{eq:LP}-\eqref{eq:LP_c3}, we have the following \cite{ParaLaR} (again ParaLaR paper):
\begin{align}
& \hspace{-8mm} \min_{x_{e,i}, \lambda_{e}} \Bigg(\sum_{i=1}^{N_{nets}}\sum_{e\in E}\left ( w_{e}+\lambda~_{e} \right )x_{e,i}-W\sum_{e\in E}\lambda~_{e}\Bigg), \label{eq:relaxed LP modified} \\
& \hspace{-8mm} \text{{Subject to}} \hspace{0.1cm} A_{i}x_{i}=b_{i}, \hspace{0.1cm} i=1,2,...,N_{nets},~\refstepcounter{equation}\subeqn \\ 
& \hspace{-8mm} \qquad\qquad x_{e,i}=0 \hspace{0.1cm}or \hspace{0.1cm} 1 \quad \text{{and}} \subeqn \\
& \hspace{-8mm} \qquad\qquad \lambda_{e}\geq 0. \label{eq:relaxed LP modified_C} \subeqn
\end{align}
In the above LP, $\left(w_{e}+\lambda_{e}\right)$ is the new cost associated with the edge \textit{e}. As earlier, this LP can be easily solved in a parallel manner.

In \eqref{eq:relaxed LP modified}--\eqref{eq:relaxed LP modified_C}, we have two sets of variables $x_{e,i}$ and $\lambda_e$. Since the decision variables $x_{e,i}$ can have values either $0$ or $1$, this LP is a Binary Integer Linear Program (BILP) that is non-differentiable \cite{JIFS_Simplex, JIFS_LPDuality, JIFS_LP}. Hence, the traditional methods such as the Simplex method~\cite{JIFS_Simplex}, the interior point method~\cite{Interior_point}, etc. fail here. The sub-gradient based methods~\cite{JIFS_Subgradient, Subgradient} are iterative methods for solving optimization problems without stringent differentiability requirements. In these methods, the variable (say \textit{x}) is updated as $x^{k+1}=x^{k}-\alpha^{k}g^{k}$, where $\alpha^k$ is the step size, $g^k$  is a~sub-gradient of the objective function, and the superscript ($k$ or $k+1$) denotes the iteration number. Since a sub-gradient based algorithm will not give binary solutions, which we need (recall $x_{e,i}$ can be $0$ or $1$), we use it to compute the Lagrange multipliers $\lambda_e$ only. For solving $x_{e,i}$, we use the minimum Steiner tree algorithm.

There are many variants of the sub-gradient based methods available such as the projected method~\cite{Subgradient}, the primal--dual method~\cite{Primal_Subgradient}, the conditional method \cite{Subgradient_Thesis}, the deflected method~\cite{Subgradient_Thesis}, etc. In our ParaLarPD algorithm~\cite{ParaLarPD}, which as earlier improved the ParaLaR algorithm~\cite{ParaLaR}, we demonstrated the superiority of using the primal--dual method with computation of the Lagrange multipliers done as below.
\begin{align}\label{eq:Lagrangian multiplier}
& \hspace{-4mm} \lambda_{e}^{k+1}=\lambda_{e}^{k}+\alpha^{k}\max\left(0,\sum_{i=1}^{N_{nets}}x_{e,i}-W\right), 
\end{align}
where $\sum_{i=1}^{N}x_{e,i}-W$ is a~sub-gradient of the objective function at the $k^{th}$ iteration--the partial derivative of the objective function in \eqref{eq:relaxed LP modified}. Also $\lambda_{e}^{0}$ is taken as zero for all edges.

In our ParaLarPD paper~\cite{ParaLarPD}, we also proposed a new step size updation strategy that works better than the corresponding technique proposed in the ParaLaR paper~\cite{ParaLaR}. That is,
\begin{align}\label{eq:step_size}
& \alpha~^{k}=\left ( 1/k \right )/\left \| T^{k} \right \|_{2},
\end{align}
where \textit{k} is the iteration number, $T^k$ is the Karush--Kuhn--Tucker (KKT) operator of the objective function \eqref{eq:relaxed LP modified}, and $\left \| T^{k} \right \|_{2}$ is the 2-norm of $T^k$.

Next, the minimum Steiner tree algorithm~\cite{SteinerTree} is used to compute $x_{e,i}$. Here, the input is a set $S$ that contains the net vertices. The intermediate goal is to compute the set of Steiner vertices for $S$, which is initially empty (say $U$). The algorithm begins by forming a triple of vertices from $S$. Next, a possible candidate Steiner vertex is found such that the total path cost from the vertices in the triple to the candidate vertex is minimized. This process is repeated for all the sets of triples to find the possible Steiner vertices, out of which $U$ is formed. Finally, the union of $S$ and $U$ is obtained using the minimum spanning tree algorithm leading to the minimum Steiner tree. The edges that are used in this tree have $x_{e,i} = 1$ and all other edges have $x_{e,i} = 0$.

After one complete iteration of the primal--dual sub-gradient algorithm as well as the Steiner tree algorithm, the value of $W$ is reduced and these steps are repeated. This help us obtain a better local minima both for the total wire length and the channel width. For easy reference the pseudo code of ParaLarPD, as published in \cite{ParaLarPD}, is given in \textbf{Algorithm \ref{algo:paralarpd}}.

\begin{algorithm*}[!h]
	\caption{ParaLarPD \cite{ParaLarPD}}
	\label{algo:paralarpd}
	\small
	\hspace*{\algorithmicindent} \textbf{Input:} Architecture description file and benchmark file. \\ 
    \hspace*{\algorithmicindent} \textbf{Output:} Route edges.
    \begin{algorithmic}[1]
		\State Run VPR with the input architecture and benchmark circuit.
		 
		\State $steiner\_points \gets \emptyset$ 
		\State $grid\_graph \gets$ InitGridGraph() 
		\State $\lambda_e = 0, \forall e\in E$		
		\For{$iter = 1$ \textbf{to} $max\_iter$}
		\State Calculate the step size $\alpha$ using \eqref{eq:step_size}. \label{line:stepsize} 
			\State $route\_edges \gets \emptyset$ 
			\algrenewcommand\algorithmicfor{\textbf{parallel\_for}}	
			\For{$i=1$ \textbf{to} $N_{nets}$}  \label{line:parallel_nets} 
				 \State $points \gets \big \{p : p \in \text{\{source and sinks of $i${th} net\}}\big \}$
					\If{$iter == 1$} 
						\State $steiner\_points[\text{$i${th}} \: net] \gets$ Min\_Span\_Tree($grid\_graph$, $points$) 
					\EndIf
					\State $route\_edges[\text{$i${th}} \,net] \gets$ Min\_Span\_Tree($grid\_graph$, $steiner\_points[\text{$i${th}} \: net] \cup points$)
			\EndFor
			\algrenewcommand\algorithmicfor{\textbf{for}}
			\While{$e$ \textbf{$\in$} $E$} \label{line:subgrad_start}
			\State Update Lagrangian relaxation multipliers $\lambda_e$ using the Equation \eqref{eq:Lagrangian multiplier}. \label{line:update_Lagrange}
			\State Update the edge weight of the $grid\_graph$ on $route\_edge$. New edge weights are $w_e+\lambda_e$. 
			\EndWhile		\label{line:subgrad_end}
			\EndFor
	\end{algorithmic}
\end{algorithm*}

\section{Proposed Approach}\label{Proposed LH}
As mentioned earlier, in our proposed work we \textit{first} perform FPGA routing using our ParaLarPD. Since some constraints are often violated by the obtained solution, \textit{second}, we develop a heuristic that converts the infeasible solution to a feasible one (i.e. tackle the issue of the constraints violation), which is discussed next.

This technique has been applied successfully in many domains \cite{Heuristic_LH1, Heuristic_LH2, Heuristic_LH3, Heuristic_LH4, Heuristic_LH5, Heuristic_LH6}. For example, \cite{Heuristic_LH1} solves a multi-plant lot-sizing problem. Here, the authors formulate an LP to minimize the production costs with the demand and the capacity constraints. The constraints are relaxed by introducing the Lagrangian multipliers. A novel Lagrangian heuristic (in the form of two feasibility stages) is applied to every solution obtained while solving for the multipliers.  The first feasibility stage consists of a local search in which the production lots are transferred amongst the time periods to ensure feasible solutions. The second feasibility stage is also a local search based strategy, however, in this, viable solutions are explored by transferring the production batches to not only the different time periods but to the different plants as well.

Another example is \cite{Heuristic_LH2} where assignment of the students to the classes (based upon their preferences) is formulated as a graph partitioning problem with the capacity constraints. This problem is further modeled as a Quadratic Program (QP), and similar to \cite{Heuristic_LH1}, the constraints are relaxed by introducing Lagrange multipliers, which are solved by the sub-gradient method. As expected, the obtained solutions are not necessarily feasible, and hence, a Lagrange heuristic is built. In this, the constraints violation are assigned probabilities based upon certain characteristics of the solution. The algorithm is again iterated with the new probability based information.

\begin{figure}[!h]
 \centering
 \includegraphics[scale=0.8]{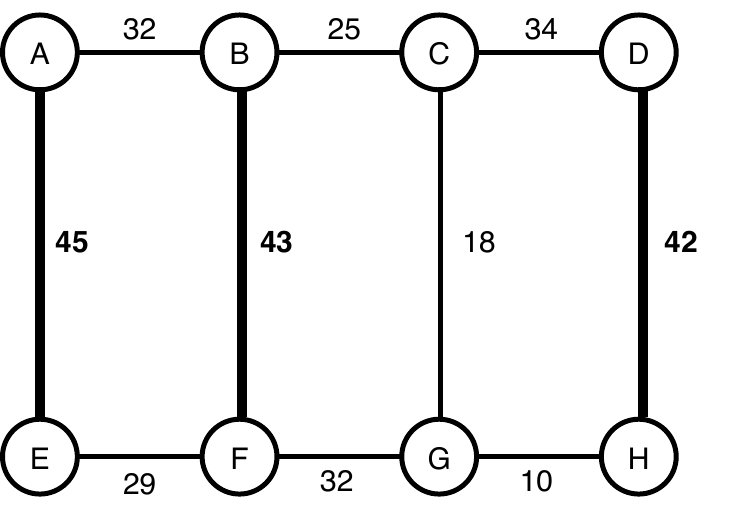}
 \caption{A sub-graph to demonstrate working of our heuristic strategy.}
 \label{fig:HeuristicExample}
\end{figure}

Our basic Lagrangian heuristic to remove the constraints violation in ParaLarPD consists of the five steps as below. Here, we initially explain these steps using the example shown in Fig. \ref{fig:HeuristicExample}, and then in the form of an algorithm. In this example, the channel widths as computed by ParaLarPD are written next to the corresponding edge in the figure. Since $W$ is taken as forty, we have three edges where the constraints violation occur. That is AE, BF, and DH that are highlighted in bold in Fig \ref{fig:HeuristicExample}. 

\begin{figure}[!h]
 \centering
 \includegraphics[scale=0.8]{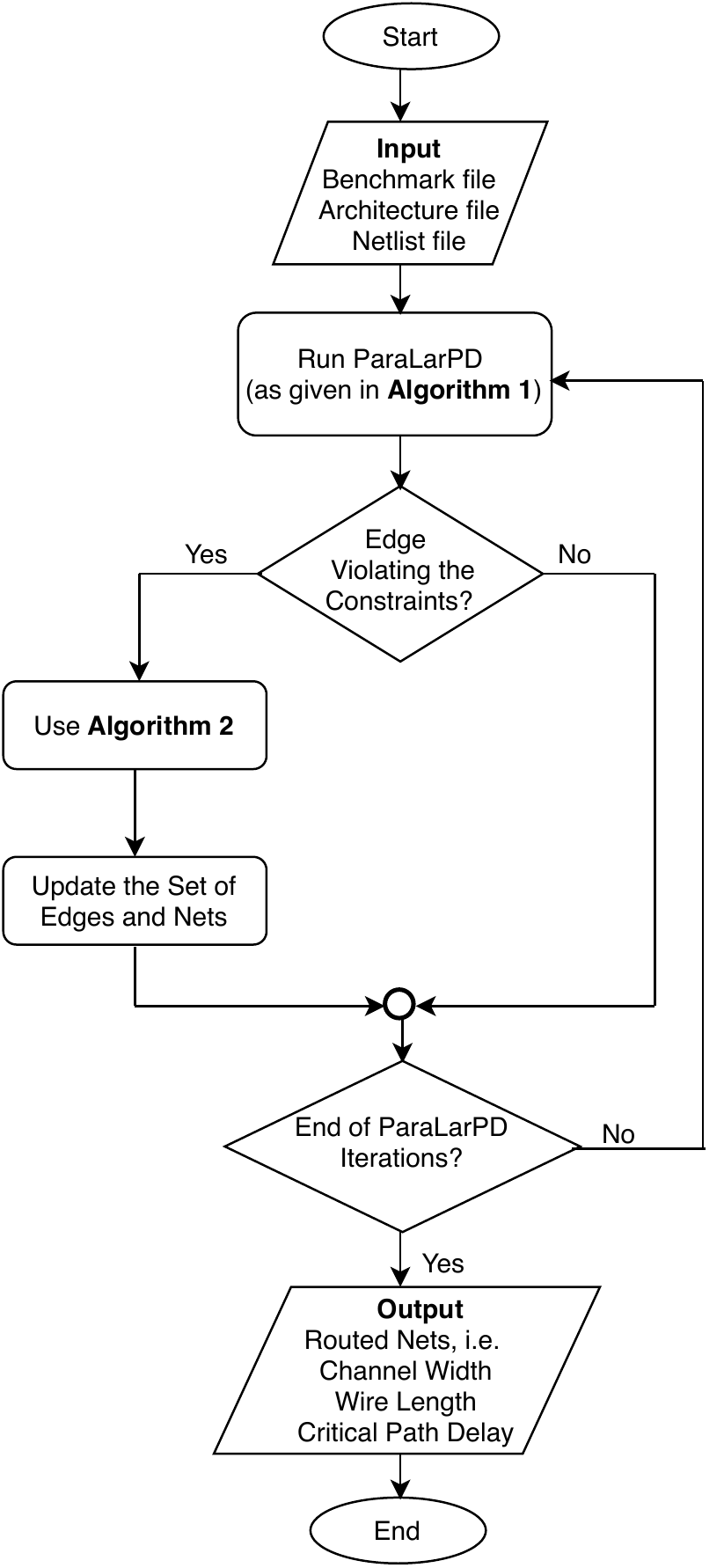}
 \caption{Data flow diagram of our ParaLarH.}
 \label{fig:DFD of ParaLarH}
\end{figure}

\begin{enumerate}
\item Pick an edge with the constraints violation, and find a new alternate path between the nodes of this edge using any path finding algorithm. There may be many alternative paths possible so pick any one. If the new path contains an edge that already has the constraints violation, then drop it and move to the next alternative path.

For example here, without loss of generality, the edge picked is BF and the first alternate path chosen is BA -> AE -> EF. Since this path contains the edge AE, which violates the constraints, and hence, we drop it and pick the next possible path (BC -> CG -> GF) where no such violation occurs. 
\item Next, compute the available capacity of each edge in the new path to route more nets without the constraints violation. Minimum of these capacities is termed as Threshold, and used further. Mathematically,  
\vspace{-5mm}
\begin{equation}
\begin{aligned}
& \text{Threshold} = {} \min(W - \sum_{i=1}^{N_{nets}}x_{e_{k},i}) \nonumber \\
&  \qquad \qquad \forall k \in \{\text{edges in the new path}\}.  \nonumber
\end{aligned}
\end{equation}

For our example, the value of Threshold is 8. 
\item Calculate the amount of violation $$d = \sum_{i=1}^{N_{nets}} x_{e,i}   - W$$ for the edge under consideration $e$. Further, calculate the number of nets where the constraints violating edge needs to be replaced by the selected new path. This is computed as
\begin{equation}
\begin{aligned}\label{eq:q}
q = \min(\text{Threshold},d)
\end{aligned}
\end{equation}
so that no edge in the added new path has the constraints violation.

For the edge under consideration (BF), $d = 43 - 40 = 3$, and hence, $q = \min(8,3) = 3$.
\item Finally, replace this edge under consideration with the selected path in $q$ number of nets.

In this example, this corresponds to replacing BF with BC -> CG -> CF in 3 nets.
\item If in \eqref{eq:q} above, $\text{Threshold} < d$, then we would have not completely eliminated the constraints violation in the edge under-consideration. In this case, the search for the alternate path needs to resumed from the start until the violation is completely eliminated or no such path exists.
\end{enumerate}
We repeat the above steps for all the edges that are violating the constraints. This violation is directly related to the minimum channel width (discussed earlier), i.e. we improve this requirement as well. \textbf{Algorithm \ref{algo:Heuristic Design}} describes our heuristic design in an algorithmic form. The points above map to the respective line numbers in the algorithm, which is termed as ParaLarH. For enhanced clarity, we describe ParaLarH via a data flow diagram as well (in Fig. \ref{fig:DFD of ParaLarH}).

\begin{algorithm}[!h] 
    \caption{Heuristic Design}
    \label{algo:Heuristic Design}
    \small
    \hspace*{\algorithmicindent} \textbf{Input:} Set of nets and edges that are being used; and the decision variables determined by ParaLarPD algorithm. \\ 
    \hspace*{\algorithmicindent} \textbf{Output:} Updated set of nets and edges that are being used.
\begin{algorithmic}
        \For{(each edge $e \in E$)} \nonumber
        
        \While{($d = \sum_{i=1}^{N_{nets}} x_{e,i} - W \ge 0 $)} \label{algo:while loop}
        \vspace*{-4mm}
	\State
	\begin{enumerate}[leftmargin=+.5in]
	\item  Find a path using any path finding algorithm $p: e_{1}e_{2}\cdots e_{r-1} e_{r}$ between the end points of the edge $e$ such that \label{algo:BFS}
\begin{align}
& e_{1}.start = e.start,  \nonumber \\ 
& e_{j}.end = e_{j+1}.start \;  \nonumber \\ 
& \qquad\qquad \forall j \in \{1,2,\hdots,r-1 \}, \nonumber \\ 
& e_{r}.end = e.end, \nonumber \\
& \sum_{i=1}^{N_{nets}}x_{e_{j},i} \leq W \; \forall j \in \{ 1,2,\hdots,r \},   \nonumber  \label{eq:algoBFS_4}
\end{align}
If there is no such alternative path available for the current constraint violating edge, then break. 

        \item Compute \label{algo:thresh}
        \begin{align}
        & \text{Threshold} = {} \min(W - \sum_{i=1}^{N_{nets}}x_{e_{k},i})  \\
        & \qquad\qquad\qquad \forall k \in \{1,2,\hdots,r\}.
	\end{align}
        \item Calculate $q = \min(\text{Threshold},d).$  \label{algo:q}

        \item If ${Net}^{e} = \{ N^{1}_{nets},N^{2}_{nets},\hdots,N^{t}_{nets}\}$ denotes the $t$ nets where edge $e$ is used. Replace $e$ with path $e_{1} e_{2} \cdots e_{r}$ in $q$ such nets. Usually $t > q$. 

\hspace{-3mm}// The Point 5 as discussed in text maps to the \texttt{while} statement above. 

	\end{enumerate}
        	 
         \EndWhile
         \EndFor 
 \end{algorithmic}
\end{algorithm}

\subsection{Other Variations of our Heuristic}\label{heuristic other Variations }
Next, we discuss some variants of ParaLarH. As mentioned earlier, these variations are designed to help reduce the constraints violation further, however, they do negligibly increase the computational cost of the overall algorithm. 
\begin{enumerate}[label=(\Alph*)]
\item The first variant is based upon the fact that there may exist multiple paths between any two end points, and in Step 1 above we should pick the one that gives the best results. Hence, instead of picking just one path randomly, we  pick $\beta$ number of paths. Further, we perform Steps 1, 2 and 3 for all these $\beta$ paths. 
\item In the second variant, in Step 4 above we begin by sorting the $t$ nets where the edge under consideration $e$ is used. This sorting is done in the increasing order of the number of new edges that get added to each net while eliminating $e$. Then, we replace $e$ with the new path in the first $q$ nets ensuring minimization of the overall constraints violation.
\end{enumerate}

The results obtained by our basic heuristic and the above variants are approximately the same. Therefore, in the next section, without the loss of generality, we present the results for the second variant. As mentioned in the Introduction, the extra computation in implementing the heuristic does increase the overall runtime of our algorithm. However, this is easily offsetted by using more threads in a parallel setting (discussed further in the next section). 

\section{Experimental Results}\label{Experimental Results}

We perform multiple experiments to demonstrate the usefulness of our ParaLarH algorithm. In Section \ref{Comparison with ParaLarPD}, we compare ParaLarH with ParaLarPD \cite{ParaLarPD}, the algorithm we improve upon, in multiple ways. In Section \ref{Comparison with Other Algorithms}, we compare ParaLarH with another algorithm of the same family (ParaLaR \cite{ParaLaR}), the current standard (VPR \cite{VPR}), and two other sparingly used algorithms (RVPack~\cite{YuanWangThesis} and GGAPack2~\cite{YuanWangThesis}).

We perform experiments on the MCNC benchmark circuits~\cite{MCNC} that range from small-sized to large-sized logic blocks. Experiments are done on a single Intel Xeon (R) CPU E5-1620 V3 machine running at 3.50 GHz with 64 GB RAM. The operating system is Ubuntu 14.04 LTS, and the kernel version is 3.13.0-100. Our codes are written in C++11 and compiled using GCC version 4.8.4 with O3 optimization flag. After compilation, the resulting codes are run using different number of threads from the Intel Threading Building Blocks (TBB) libraries.

\subsection{Comparison with ParaLarPD}\label{Comparison with ParaLarPD}

Here, \textit{first}, we discuss the setup of our experiments. \textit{Second}, we compare ParaLarH with ParaLarPD using the main metric of the constraints violation as well as the related metric of the minimum channel width. Here, we also compare the two algorithms using the metrics of the total wire length and the critical path delay. \textit{Third}, and finally, we compare ParaLarH with ParaLarPD using the speedups obtained because of multithreading.

\subsubsection{Setup}\label{Setup}

We use the most common architecture parameters \cite{VPR, Deterministic_FPGA, YuanWangThesis, ParaLarPD} as given in Table \ref{tab1}. Here, the value of $N$ specifies that the CLBs in the architecture contains ten Fracturable Logic Elements (FLEs). The value of $K$ specifies that each FLE has six inputs. The values of $F_{cin}$ and $F_{cout}$ specify that every input and output pin is driven by 15\% and 10\% of the tracks in a channel, respectively. We also perform experiments with $F_{cin}=1$ and $F_{cout}=1$. In this paper, we do not report these results because in most of the modern FPGA designs, input and output pins are not driven by 100\% of the tracks in a channel. However, our ParaLarH gives better results than ParaLarPD for this as well. The value of $F_s$ specifies the number of wire segments that can be connected to each wire where the horizontal and the vertical channels intersect. This value can only be a multiple of $3$. In this paper, we report the results for $F_s = 3$. We also perform experiments with $F_s=6$, the results of which are not reported here. However, for this case too, ParaLarH gives better results than ParaLarPD. The value of Length specifies the number of logic blocks spanned by each segment. Here, we take this value as 4, though our proposed FPGA routing can be used for architectures with other values of Lengths as well, e.g. Length $= 1$ or a mix of Length $=1$ and Length $=4$. 

\begin{table}[!h]
\caption{FPGA design architecture parameters used in our experiments.}\label{tab1}
\centering
\small	
\setlength{\tabcolsep}{10pt}
{\begin{tabular}{llllll}\toprule
N & K & $F_{cin}$ & $F_{cout}$ & $F_s$ & Length \\
\midrule
10 & 6 & 0.15 & 0.10 & 3  & 4 \\
\bottomrule
\end{tabular}}{}
\end{table}

Initially, the circuits are packed and placed using VPR. After that, routing is performed by the respective algorithm. There is no general rule of choosing the initial value of the channel width for experimental purposes. However, a value of 20\% to 40\% more than the minimum channel width obtained from VPR is commonly used \cite{ParaLaR, ParaLarPD, Deterministic_FPGA}. Our algorithms are initialized with the initial channel width ($W$) as $1.2W_{min}$, where $W_{min}$ is the minimum channel width obtained from VPR. We also perform experiments with an initial $W$ as $1.4W_{min}$, which does not change the results. We use an upper limit of $50$ for the number of iterations and perform 100 runs of all the algorithms. The best results out of these are reported.

\subsubsection{Basic Comparison}\label{Basic Comparison}

The results here are independent of the number of threads used, and hence, we give results for a single thread only. We use up to five significant digits to report our data. Specifically, we round the constraints violation, the minimum channel width and the critical path delay to two decimal places, and the total wire length to the nearest integer. Here, we also report the geometric mean (Geo. Mean) of all the values obtained for the different benchmark circuits, which indicates the central tendency of a set of numbers and is commonly used~\cite{ParaLaR, ParaLarPD}.

In Table \ref{tab2}, we compare the constraints violation and the minimum channel width of ParaLarH and ParaLarPD. As evident from this table, for all the benchmarks, ParaLarH substantially improves both these metrics. The average constraints violation reduces to half, and the average reduction in the minimum channel width is 10\%, which as per the FPGA routing domain experts is considered very good.



\begin{table}[!h]
\caption{Comparison of the constraint violation and the minimum channel width between our proposed ParaLarH and ParaLarPD~\cite{ParaLarPD}.}\label{tab2}
\centering
\small	
\setlength{\tabcolsep}{1.8pt}
{
\begin{tabular}{lrrrrr}
\hline
\multicolumn{1}{l}{\begin{tabular}[c]{@{}l@{}}Benchmark\\ Circuits\end{tabular}} & \multicolumn{2}{c}{\begin{tabular}[c]{@{}c@{}}Absolute Constraint\\ Violation\end{tabular}} & \multicolumn{2}{c}{\begin{tabular}[c]{@{}c@{}}Minimum Channel\\ Width\end{tabular}} \\ \cline{2-5}

  & \multicolumn{1}{c}{\textbf{ParaLarH}} & \multicolumn{1}{c}{ParaLarPD} & \multicolumn{1}{c}{\textbf{ParaLarH}} & \multicolumn{1}{c}{ParaLarPD} \\
\hline
Alu4      & \textbf{4.67}                         & 5.54                          & \textbf{34.67}                        & 35.54                         \\
Apex2     & \textbf{5.12}                         & 10.08                         & \textbf{45.12}                        & 50.08                         \\
Apex4     & \textbf{1.87}                         & 5.58                          & \textbf{41.87}                        & 45.58                         \\
Bigkey    & \textbf{5.18}                         & 9.04                          & \textbf{15.18}                        & 19.04                         \\
Clma      & \textbf{5.32}                         & 16.44                         & \textbf{70.32}                        & 81.44                         \\
Des       & \textbf{5.40}                          & 11.17                         & \textbf{25.40}                         & 31.17                         \\
Diffeq    & \textbf{2.67}                         & 7.52                          & \textbf{32.67}                        & 37.52                         \\
Dsip      & \textbf{4.86}                         & 5.63                          & \textbf{24.86}                        & 25.63                         \\
Elliptic  & \textbf{5.20}                          & 12.42                         & \textbf{50.20}                         & 57.42                         \\
Ex1010    & \textbf{4.78}                         & 8.83                          & \textbf{44.78}                     & 48.83                         \\
Ex5p      & \textbf{4.23}                         & 8.31                          & \textbf{59.23}                        & 63.31                         \\
Frisc     & \textbf{6.45}                         & 13.71                         & \textbf{61.45}                        & 68.71                         \\
Misex     & \textbf{4.56}                         & 7.27                          & \textbf{39.56}                        & 42.27                         \\
Pdc       & \textbf{4.34}                         & 8.67                          & \textbf{69.34}                   	     & 73.67                         \\
S298      & \textbf{4.87}                         & 9.00                             & \textbf{34.87}                     & 39.00                            \\
S38417    & \textbf{5.34}                         & 10.48                         & \textbf{45.34}                     & 50.48                         \\
Seq       & \textbf{4.68}                         & 8.85                          & \textbf{44.68}                        & 48.85                         \\
Spla      & \textbf{4.92}                         & 9.41                          & \textbf{54.92}                        & 59.41                         \\
Tseng     & \textbf{4.87}                         & 9.65                          & \textbf{34.87}                        & 39.65                         \\ \hline
Geo. Mean & \textbf{4.56}                         & 8.98                         & \textbf{40.99}                        & 45.55                         \\ 
\hline
\end{tabular}
}
\end{table}

While this work focuses on minimizing the constraints violation, we measure our algorithm\textquotesingle s impact on other metrics as well, i.e., the total wire length and the critical path delay. In Table \ref{tab3}, we compare these metrics for ParaLarH and ParaLarPD. We observe almost a negligible increment in the total wire length obtained by ParaLarH as compared to ParaLarPD (on an average $0.92\%$). If we look at the easily measurable critical path delay, we observe that on an average critical path delay of ParaLarH is $7.60\%$ less as compared to ParaLarPD, which again is considered a very good improvement.

\begin{table}[!h]
\caption{Comparison of the total wire length and the critical path delay between our proposed ParaLarH and ParaLarPD~\cite{ParaLarPD}.}\label{tab3}
\centering
\small
\setlength{\tabcolsep}{1.8pt}
{
\begin{tabular}{lrrrrr}
\hline
\multicolumn{1}{l}{\begin{tabular}[c]{@{}l@{}}Benchmark\\ Circuits\end{tabular}} & \multicolumn{2}{c}{\begin{tabular}[c]{@{}c@{}}Total Wire Length\end{tabular}} & \multicolumn{2}{c}{\begin{tabular}[c]{@{}c@{}}Critical Path Delay\end{tabular}} \\ \cline{2-5}

  & \multicolumn{1}{c}{\textbf{ParaLarH}} & \multicolumn{1}{c}{ParaLarPD} & \multicolumn{1}{c}{\textbf{ParaLarH}} & \multicolumn{1}{c}{ParaLarPD} \\ \hline
Alu4                               & \textbf{5030}        & 5030            & \textbf{6.99}    & 7.30    \\
Apex2                             & \textbf{7978}        & 7935            & \textbf{7.30}   & 7.41    \\
Apex4                             & \textbf{5807}        & 5630            & \textbf{6.51}    & 7.08    \\
Bigkey                            & \textbf{3927}        & 3896            & \textbf{3.32}    & 4.01     \\
Clma                             & \textbf{49474}       & 49278           & \textbf{15.42}   & 15.46    \\
Des                                & \textbf{7043}        & 6952            & \textbf{5.47}    & 5.55    \\
Diffeq                             & \textbf{4693}        & 4349            & \textbf{5.84}    & 5.65     \\
Dsip                               & \textbf{4771}        & 4778            & \textbf{3.19}   & 3.62     \\
Elliptic                           & \textbf{15253}       & 15125           & \textbf{7.53}  & 10.83    \\
Ex5p                              & \textbf{4916}        & 4889            & \textbf{6.32}    & 6.94    \\
Ex1010                        & \textbf{23603}       & 23596           & \textbf{12.00}  & 14.57    \\
Frisc                             & \textbf{19713}       & 19484           & \textbf{12.68}  & 13.13    \\
Misex3                          & \textbf{5195}        & 5194            & \textbf{6.19}  & 6.49    \\
Pdc                              & \textbf{30435}       & 30423           & \textbf{12.39}  & 12.63    \\
S298                             & \textbf{5256}        & 5250            & \textbf{11.67}    & 12.71    \\
S38417                             & \textbf{21962}  & 21907          & \textbf{9.11}  & 10.44    \\
Seq                                 & \textbf{7685}        & 7654           & \textbf{6.00}  & 6.14    \\
Spla                               & \textbf{20139}       & 20117           & \textbf{10.23}   & 10.43    \\
Tseng                             & \textbf{2491}        & 2484            & \textbf{5.78}   & 5.78     \\
\hline
Geo Mean                       & \textbf{9124.20}     & 9041.01         & \textbf{7.42} & 8.03 \\
\hline
\end{tabular}
}
\end{table}

\subsubsection{Comparison of Speedups}\label{Speedup Comparison}

\begin{figure*}[]
	\centering
	\includegraphics[scale=0.45]{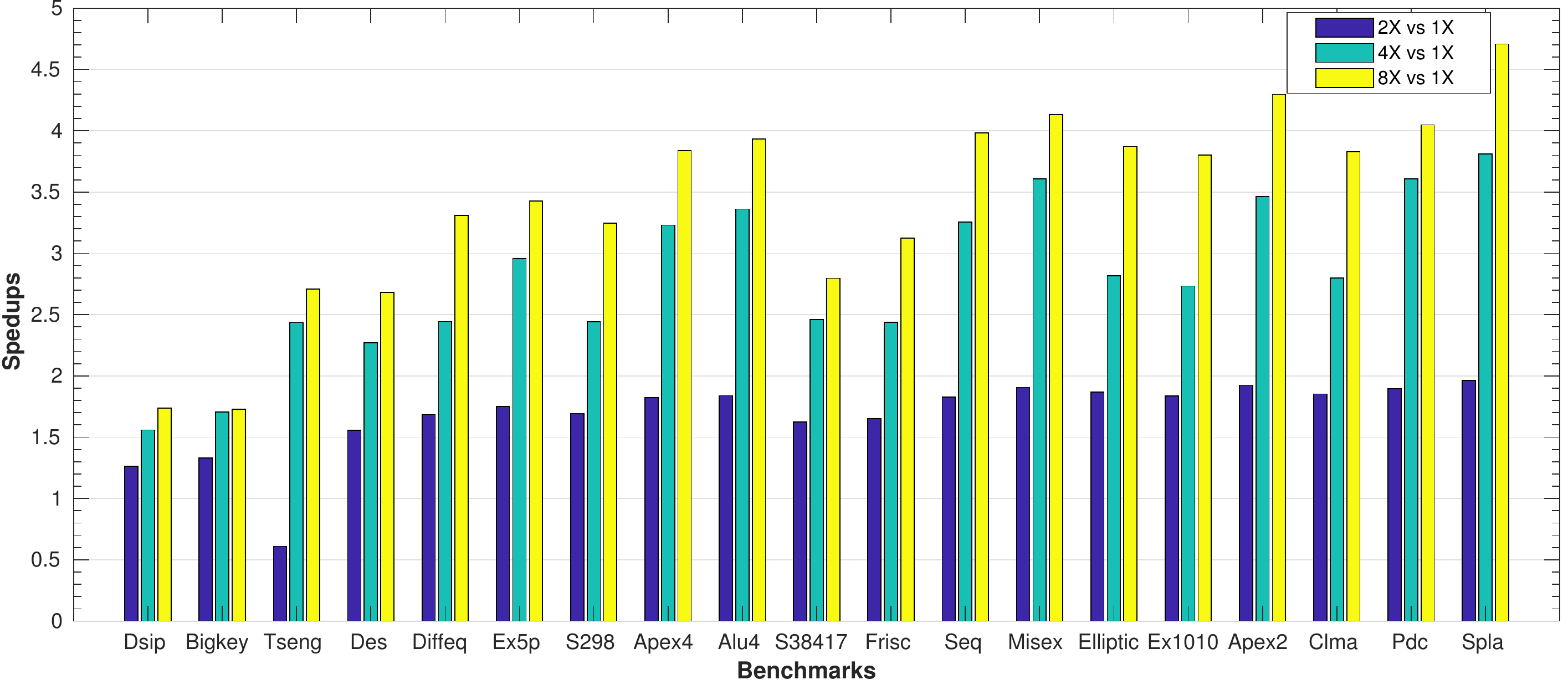}
	\caption{Speedups of each benchmark using ParaLarH when running it with $2$, $4$ and $8$ threads.}
	\label{fig:speedups graph}
\end{figure*}

The speedups obtained in executing the benchmarks via ParaLarH in a parallel setting are given as a bar graph in Fig. \ref{fig:speedups graph}. In this figure, on the x-axis we have the benchmark circuits arranged in the increasing order of their execution time when running them with one thread. This time is directly proportional to the benchmark size, and arranging them this way ensures that the benchmarks are sorted in the order of their increasing size (used below). On the y-axis, we have the speedups in execution of these benchmark circuits when using $2$ threads, $4$ threads, and $8$ threads in ParaLarH. 

We make two observations from this bar graph. \textit{First}, the benchmarks of larger sizes (towards the right of the graph) have more speedups as compared to those of smaller sizes (towards the left of the graph). \textit{Second}, on an average, 2, 4, and 8 threads give speedups of 1.63, 2.74, and 3.32, respectively.

In ParaLarPD \cite{ParaLarPD}, the average speedups when using 2, 4, and 8 threads comes to be 1.80, 3.11, and 5.11, respectively. If we compare the speedups obtained from ParaLarH with ParaLarPD, we observe that there is a slight deterioration in the case of ParaLarH.

This drop in speedups is acceptable because of two reasons. \textit{First}, as compared to ParaLarPD, ParaLarH substantially improves the constraints violation (halving it), the minimum channel width (reducing it by 10\%), and the critical path delay (reducing it by 8\%) while keeping the total wire length almost the same. \textit{Second}, this slight loss in speedups is easily compensated by using more number of threads.

\subsection{Comparison with the Other Algorithms}\label{Comparison with Other Algorithms}

As mentioned earlier, here we compare ParaLarH with the other algorithms for FPGA routing. That is, ParaLaR, VPR, RVPack and GGAPAck2. This comparison is done using all the above discussed metrics; the minimum channel width\footnote{This captures the constraints violation implicitly.}, the total wire length, the critical path delay, and the speedups. 

For ParaLaR and VPR, we use the setup as discussed in Section \ref{Setup}. RVPack and GGAPack2 design is fundamentally different from the other algorithms (i.e. ParaLarH, ParaLarPD, ParaLaR, and VPR), and hence, the above mentioned setup cannot be used. Further, replicating the setup of RVPack and GGAPack2 is challenging as well due to the involved randomness. Hence, we use VPR as the base algorithm against which we compare ParaLarH, ParaLaR, RVPack and GGAPack2 with experiments done here for the first two algorithms and data picked from the original paper (\hspace{1sp}\cite{YuanWangThesis}) for the last two. Since in VPR, RVPack, and GGAPack2 there is no concept of parallelization, all comparisons are done using a single thread. 
 
This comparison is given in Table \ref{tab5}, with the best improvements highlighted in bold. Here, a negative sign indicates deterioration. From this table, we can see that for the percentage gain in the minimum channel width as well as the critical path delay, ParaLarH triumphs the other three algorithms. Its percentage improvement in the total wire length is also almost the best (very close to that of ParaLaR). Regarding speedups, the percentage gain in ParaLarH is the second best (and slightly below that of ParaLaR), which is acceptable since it can be easily compensated as discussed in Section \ref{Speedup Comparison}.

\begin{table}[!h]
\caption{Performance comparison of ParaLarH with ParaLaR~\cite{ParaLaR}, VPR \cite{VPR}, RVPack \cite{YuanWangThesis} and GGAPack2 \cite{YuanWangThesis}. \label{tab5}}
\centering
\small
\setlength{\tabcolsep}{1.8pt}	
\renewcommand{\arraystretch}{1}	
{
\begin{tabular}{lcccc}
\hline
\multirow{2}{*}{Algorithms} & \multicolumn{4}{c}{$\%$ Improvement over VPR}  \\ \cline{2-5}
 & \begin{tabular}[c]{@{}c@{}}Minimum \\ Channel Width\end{tabular} & \begin{tabular}[c]{@{}c@{}}Total Wire\\ Length\end{tabular} & \begin{tabular}[c]{@{}c@{}}Critical\\ Path Delay\end{tabular} & \multicolumn{1}{c}{Speedup}       \\ \hline
ParaLarH & \textbf{34.68} & 48.39 & \textbf{9.95}	& 2.28 				\\ 
ParaLaR & 9.08 & \textbf{48.83} & 4.00 & \textbf{2.70}				\\ 
RVPack       & 12.37 & 7.19  & -31.64 & 1.28            \\ 
GGAPack2     & 2.23  & -5.78 & -35.94 & \textless{}-100 \\ \hline
\end{tabular}}
\end{table}

\section{Conclusions and Future Work}\label{Conclusion}
In this work, we improve upon ParaLarPD, the best available algorithm for the parallel routing of Field Programmable Gate Array (FPGA) design flow. ParaLarPD formulates the design process as a Linear Program (LP) and solves it using the Lagrange relaxation, the sub-gradient method, and the Steiner algorithm. Here, we improve this algorithm\textquotesingle s Lagrange relaxation process by introducing a set of Lagrange heuristics resulting in substantial reduction of the constraints violation by the solution vector. We term our algorithm as ParaLarH.

With experiments on the MCNC benchmark circuits we show that as compared to ParaLarPD, on an average, ParaLarH halves the constraints violation, reduces the minimum channel width metric by 10\% and the easily measurable metric of critical path delay by 8\%. The extra work in implementing the heuristic does slightly deteriorate the parallelization speedups of ParaLarH as compared to that of ParaLarPD, however, this is easily fixable by using more number of threads. In future, we plan to work towards designing algorithms that would completely remove the constraints violation. We also plan to apply our techniques to the field of Internet of Things (IoT) where similar design challenges arise.



\end{document}